\documentclass[conference, 9pt]{IEEEtran}
\IEEEoverridecommandlockouts
\usepackage{cite}
\usepackage{amsmath,amssymb,amsfonts}
\usepackage{multirow}
\usepackage{graphicx}
\usepackage{textcomp}
\usepackage[table,xcdraw]{xcolor}
\usepackage{listings}
\usepackage{glossaries}
\usepackage{enumitem}

\usepackage{latexsym}
\usepackage{algorithm}
\usepackage{algpseudocode}
\usepackage{microtype}
\usepackage{pgfplots}
\usepackage{listings}
\def\BibTeX{{\rm B\kern-.05em{\sc i\kern-.025em b}\kern-.08em
    T\kern-.1667em\lower.7ex\hbox{E}\kern-.125emX}}
\definecolor{vgreen}{RGB}{104,180,104}
\definecolor{vblue}{RGB}{49,49,255}
\definecolor{vorange}{RGB}{255,143,102}

\definecolor{comment}{rgb}{0,.6,0}
\definecolor{keyword}{rgb}{.63,0,.42}
\definecolor{kw2}{rgb}{.50,.50,.15}
\definecolor{kw3}{rgb}{.42,.42,.63}
\definecolor{string}{rgb}{1,0,0}
\newacronym{llm}{LLM}{Large Language Model}
\lstdefinestyle{vcode}{
	language=Verilog,
    commentstyle=\color{comment},
	stringstyle=\color{string},
	keywordstyle=\bfseries\color{keyword},
	basicstyle=\scriptsize\ttfamily,  
    captionpos=b,
    numbers=none,	
    frame=single,
	showstringspaces=false,
	tabsize=2
}

\lstdefinestyle{vcode1}{
	language=Verilog,
    commentstyle=\color{comment},
	stringstyle=\color{string},
	keywordstyle=\bfseries\color{keyword},
	basicstyle=\footnotesize\ttfamily,
        captionpos=b,
	showstringspaces=false,
	tabsize=2,
        framesep=0pt,                     
        xleftmargin=0pt,                  
        xrightmargin=0pt,                 
}
\usepackage{comment}
\usepackage{ulem}
\newcommand{\projectname}{\textsc{LLM4SecHW}}
\newif\ifrev
\revtrue
\ifrev
  \newcommand{\xiaolong}[1]{{\color{red} [Xiaolong: #1]}}
\else  
  \newcommand{\xiaolong}[1]{}
\fi

\newif\ifrev
\revtrue
\ifrev
  \newcommand{\kaichen}[1]{{\color{green} [Kaichen: #1]}}
\else  
  \newcommand{\kaichen}[1]{}
\fi

\newif\ifrev
\revtrue
\ifrev
  \newcommand{\raj}[1]{{\color{blue} [Raj: #1]}}
\else  
  \newcommand{\raj}[1]{}
\fi

\newif\ifrev
\revtrue
\ifrev
  \newcommand{\weimin}[1]{{\color{cyan} [Weimin: #1]}} 
\else  
  \newcommand{\weimin}[1]{}
\fi
\pgfplotsset{compat=1.18}
\begin{document}

\title{LLM4SecHW: Leveraging Domain-Specific Large Language Model for Hardware Debugging\\
\thanks{Portions of this work were supported by the National Science Foundation (CCF-2019310, First Award Program of ARISE in EPSCoR 2148878).}
}

\author{

\IEEEauthorblockN{Weimin Fu\IEEEauthorrefmark{1},
Kaichen Yang\IEEEauthorrefmark{2}, 
Raj Gautam Dutta\IEEEauthorrefmark{3}, 
Xiaolong Guo\IEEEauthorrefmark{1}
and Gang Qu\IEEEauthorrefmark{4}\\
\IEEEauthorblockA{\IEEEauthorrefmark{1}Kansas State University, \IEEEauthorrefmark{2}Michigan Technological University,
\IEEEauthorrefmark{3}Silicon Assurance,
\IEEEauthorrefmark{4}University of Maryland}
}
weiminf@ksu.edu, kaicheny@mtu.edu, rajgautamdutta@siliconassurance.com, guoxiaolong@ksu.edu, gangqu@umd.edu\\
}

\IEEEoverridecommandlockouts
\IEEEpubid{\makebox[\columnwidth]{979-8-3503-4099-0/23\$31.00~\copyright2023 IEEE \hfill} \hspace{\columnsep}\makebox[\columnwidth]{ }}
\maketitle
\IEEEpubidadjcol
\begin{abstract}
This paper presents \projectname{}, a novel framework for hardware debugging that leverages domain-specific \gls{llm}. Despite the success of \gls{llm}s in automating various software development tasks, their application in the hardware security domain has been limited due to the constraints of commercial \gls{llm}s and the scarcity of domain-specific data. To address these challenges, we propose a unique approach to compile a dataset of open-source hardware design defects and their remediation steps, utilizing version control data. This dataset provides a substantial foundation for training machine learning models for hardware. \projectname{} employs fine-tuning of medium-sized \gls{llm}s based on this dataset, enabling the identification and rectification of bugs in hardware designs. This pioneering approach offers a reference workflow for the application of fine-tuning domain-specific \gls{llm}s in other research areas. We evaluate the performance of our proposed system on various open-source hardware designs, demonstrating its efficacy in accurately identifying and correcting defects. Our work brings a new perspective on automating the quality control process in hardware design. 
\end{abstract}
\glsreset{llm}

\section{Introduction}

Designing a modern hardware is becoming increasingly challenging due to the complexity of chips for applications such as IoT, AI, and Quantum Computing~\cite{ray2017system}. These intricate hardware designs are hard to test and verify, raising the risk of hidden bugs and vulnerabilities. One major reason is that existing verification and testing approaches often require the manual creation of assertions, data models, and test vectors ~\cite{aftabjahani2021special}. Furthermore, some vulnerabilities may not affect all functionalities of design, making sole reliance on functional verification insufficient for ensuring system robustness and reliability~\cite{king2013practical, xiong2021survey}.
Considering that flaws in hardware design can be primary sources of potential security vulnerabilities, it is essential to automatically identify and fix hardware bugs with minimal human intervention during the design phase.
Fine-tuning \gls{llm}s for domain-specific tasks has seen successes in fields like medicine~\cite{singhal2023towards} and software design~\cite{github2021copilot}. \gls{llm}s have been at the forefront of advancements in numerous software programming-related tasks, demonstrating their potential in automating tasks like auto code completion, malware detection, and code refactoring \cite{openai2021chatgpt}.

However, when we delve deeper into the domain of hardware security, the application of \gls{llm}s appears to be minimal or largely restricted to the use of prompts \cite{cosler2023nl2spec, sun2023towards, kande2023llm}. The use of these prompts presents several disadvantages: 1) Performance is constrained by the general  \gls{llm}s, which are designed for general use rather than domain-specific research; 2) There is a high dependency on a specific platform; 3) Privacy concerns arise \cite{gurman2023samsung}; and 4) The cost can be substantial. Overcoming these challenges is possible through fine-tuning the \gls{llm}s for domain-specific scientific areas. 
However, in the area of hardware security, the data necessary for effective fine-tuning is limited~\cite{jiang2021delving}. 
This data scarcity becomes a considerable challenge when leveraging \gls{llm}s for removing hardware flaws, especially considering that flaws in hardware design can be primary sources of potential security vulnerabilities. Furthermore, accurately identifying and localizing these bugs are paramount to prevent potential hardware design failures.

This paper introduces a \gls{llm}-based hardware debugging framework, \projectname{}, designed to address the aforementioned issues. It aims to identify bugs and provide debugging suggestions during the hardware design iteration process. Specifically, we develop an innovative data collection and preprocessing method to harness version control information from open-source hardware projects. From this information, we construct a hardware debugging-oriented dataset by filtering and processing the version control data, which is subsequently utilized to fine-tune our model. Leveraging this dataset, we fine-tune a suite of hardware domain-specific language models capable of reading hardware designs and autonomously locating and rectifying bugs. The principal contributions of this paper include:

\begin{itemize}[left=0pt]
    \item  We propose a novel approach to compile a unique dataset of open-source hardware design defects and their remediation steps, utilizing the version control data. This dataset addresses the scarcity of functional hardware data and provides a substantial foundation for training machine learning models.


    \item  \projectname{} employs fine-tuning of 7 billion parameters \gls{llm}s based on the constructed dataset, enabling the identification and rectification of bugs in hardware designs. This framework represents a pioneering approach in the application of \gls{llm}s for automated hardware bug detection and rectification. Furthermore, it offers a referable workflow for the practical application of fine-tuning domain-specific \gls{llm}s in other research fields.

    \item  We evaluate the methods' performance on various open-source hardware designs, demonstrating their efficacy in accurately identifying and correcting defects. Our solution provides a new perspective on automating the quality control process in hardware design. We will release the dataset to the public in the future.

\end{itemize}


\section{Background and Related Works}

\subsection{\gls{llm} and Code Analysis}
The sequence models pivotal in facilitating the advancement of LLMs are the Transformers \cite{vaswani2017attention}. In recent years, transformer-based models have emerged as the principal technology in predicting text-based information. Beyond the significant success in processing and generating texts and codes, LLMs are also a promising method in code analysis, especially in checking bugs and vulnerabilities. 
LLMs are also adopted in the hardware domain to facilitate code checking. In the task of generating security assertions, existing works focus on applying prompts engineering \cite{kande2023llm, nair2023generating, thakur2023benchmarking, ahmad2023fixing} to generate secure hardware code or fulfill hardware code completion. 
However, existing works on hardware code analysis suffer from insufficient design code with explicit nature language description, especially in the hardware security domain, where dedicated datasets of hardware security are sparse. The Metrics4ML\cite{10.1145/3569052.3578926} project is an excellent attempt that provides datasets to bridge the gap between industry and academia, though the work at the hardware design level remains ongoing. Shailja compiled a dataset from open-source hardware designs on GitHub\cite{https://doi.org/10.48550/arxiv.2212.11140}, but it lacks functional and debugging descriptions.



\subsection{Version Control, Git and GitHub}
In data management, version control plays a crucial role in transient and fluctuating data. Among the available tools, Git has emerged as a predominant choice for overseeing code- and text-based content~\cite{loeliger2012version}. GitHub, as an augmentation of the Git version control system, offers an online interface for developers to collaborate on and contribute to projects. It boasts numerous features, such as Commits, Pull Requests (PRs), and Issues, streamlining code versioning, review, and collaboration.

A commit in Git delineates changes made to the files within a repository. Each commit possesses a distinct identifier, typically a hash, and is accompanied by metadata detailing the author, date, and message elucidating the rationale behind the change. A commit message succinctly conveys the purpose of the modification, its justification, and potential implications. PRs, on the other hand, enable developers to propose code alterations for integration into another branch. A typical PR encompasses a title, description, multiple commit details intended for merging into the main project, and discussions within the team. Should a PR aim to address a specific challenge or task, it frequently associates with a corresponding Issue.
Issues function to monitor and manage bugs, feature enhancements, tasks, and other pertinent concerns in a project. Often initiated by users, these concerns are articulated through feedback. Components of an issue include tags like \textit{``bug,'' ``enhancement,''} and \textit{``help wanted,''} assisting team members in swiftly pinpointing and addressing concerns.

\section{Methodology}\label{sec:method}
\begin{figure*}[t]
    \centering
    \includegraphics[width=\linewidth]{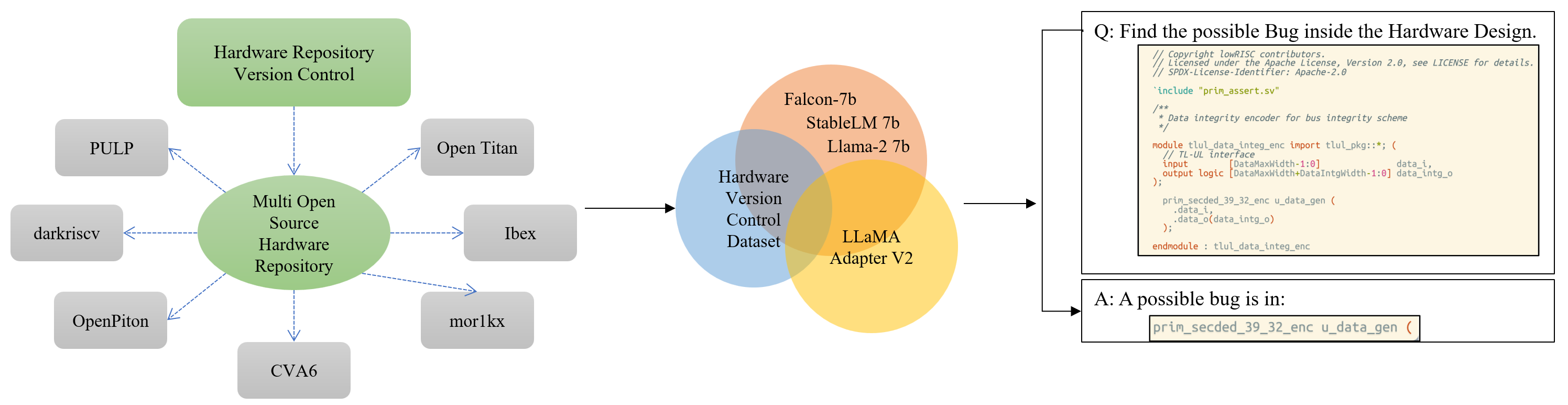}
    \caption{Method overview of the proposed \projectname{}}
    \label{fig:methodology}
\end{figure*}

\subsection{Overview of the Proposed Methodology}
The architecture of \projectname{} is illustrated in Fig. \ref{fig:methodology}. This structure is predominantly segmented into three core components, represented by the three differently colored circles in the center of the figure. On the left, the data collection process is depicted. We amass a dataset pertinent to bug-related hardware design for fine-tuning purposes through version control information from multiple open-source hardware projects.  The blue circle of the Figure 1, represents hardware debugging dataset obtained after rigorous filtering, processing, and enhancement. The orange circle of the figure denotes the three models we choose for our study, while the yellow circle encapsulates the fine-tuning methodologies. Upon completion of the fine-tuning process, the refined models are equipped to interpret the input hardware design and a concise task prompt to produce a refined or corrected version.


\subsection{Data Gathering, Clean, and Enhancement}

This section elaborates on the data collection and processing, training LLMs to understand and rectify potential flaws in hw design. 
\subsubsection{Version Control Information of Hardware Design}
We begin by assembling a collection of notable open-source hardware designs from GitHub, including CVA6~\cite{zaruba2019cost}, CVA5~\cite{TAIGA}, OpenTitan~\cite{johnson2018titan}, Ibex~\cite{schiavone2017slow}, mor1kx~\cite{mor1kx_git}, OpenPiton~\cite{balkind_openpiton_2016}, PULP~\cite{8715500}, and darkriscv~\cite{darklife2023darkriscv}, among others. Having curated this list, we then utilizes the GitHub REST API to retrieve commit, issue, and PR details from the associated repositories.
Within the amassed data, two distinct filtering phases were undertaken. First, PRs unrelated to hardware design and their corresponding commits were screened based on PR labels. Subsequent refinement targeted modifications in commits, eliminating those unrelated to hardware design based on file type. The curated set extracted pre-fix hardware design code, potentially harboring bugs, and post-fix hardware design code, considered bug-free. Additionally, commit messages, PR descriptions, and issue content were captured. We constructed a raw dataset containing over $11,000$ hardware design files, each with pre- and post-correction versions.

\subsubsection{Data Clean}
However, the raw dataset is unsuitable for fine-tuning LLMs directly.Two main challenges arise:
\begin{itemize}[left = 0pt]
    \item Data Repetitiveness: Repetitive source data, a commonplace in GitHub projects, could pose considerable adverse biases during LLM training~\cite{allamanis2019adverse}. In response, we instituted a filtration process to expunge redundant entries.
    \item Context Length Limitations: LLMs adhere to specific context length constraints. Maintaining input compatibility with LLMs requires the removal of lengthy files that exceed these set limits. Although the models' tokenizers used in \projectname{} have distinct implementations, their foundational mechanism is the same. Among the models we considered, Falcon 7B has the shortest allowable context length, roughly half that of the others. Consequently, we used Falcon 7B as our benchmark for segmentation.
\end{itemize}

\noindent Concurrently, files containing less than $15$ tokens, about ten words, were determined to be overly succinct to convey meaningful information. We excluded these files to enhance the overall information density of the dataset. The dataset encompasses over $3,000$ file pairs, which, when tokenized using the Falcon tokenizer, amounts to approximately $6$ million tokens.

\subsubsection{Data Enhancement for Downstream Tasks} 
Two downstream tasks are included in \projectname{} -- bug localization and bug repair. Our method accurately detect defects within original designs, thereby facilitating bug localization. Simultaneously, it can establish a relationship between hardware design buggy versions and corresponding bug-free versions to train LLMs for hardware bug repair.
To improve the support for both downstream tasks, \projectname{} enhances the dataset from the following aspects. 

\begin{itemize}[left=0pt]
    \item Linguistic and knowledge for hardware design: The dataset includes original and preprocessed code pairs via Verilator~\cite{snyder2013verilator} and correlation between code and finite-state automata to enhance the model's understanding of hardware design principles and language grammar.
    \item Knowledge of hardware bugs: The data includes specific commits, issues, and PR pairs to enhance the model's understanding of the relationship between hardware design programming language defects and issues described in natural language.    
\end{itemize}



\noindent The primary challenge during data enhancement arises from the version control information offered by open-source hardware communities. The information often comes with limited documentation and needs more standardized formats. We crafted comprehensive commit messages enriched with details by leveraging the commit message and pre- and post-revision code.
The dataset consists of $15,000$ samples, amounting to approximately $28$ million tokens.
The dataset is divided into training ($75\%$), validation ($15\%$), and testing ($10\%$) sets.


\subsection{Fine-Tuning Domain-Specific LLMs}
This section elaborates on the models chosen for \projectname{}, the rationale behind their selection, and the fine-tuning method. 
Table \ref{tab:model} presents a comprehensive summary of the employed models, detailing aspects such as quantity of parameters, layer counts, hidden unit dimensions, and content lengths, in addition to fine-tuning hyperparameters like the learning rate, $\beta_1$, and $\beta_2$ for the AdamW optimizer~\cite{loshchilov2019decoupled}.


\subsubsection{Model Choices for Fine-Tuning}
We selected three models. First, the StableLM model~\cite{gpt-neox-library} from the GPT-NeoX series~\cite{gpt-neox-library}, which gained prominence alongside ChatGPT in late 2022, offering parameter sizes of 3b and 7b. Following that, the Falcon model~\cite{falcon40b}, which topped the HuggingFace Open LLM Leaderboard in July 2023\cite{Open-LLM-Leaderboard-Report-2023}, is available in 7b and 40b sizes. Lastly, the LLama2 model~\cite{touvron2023llama}, which emerged in August 2023 and rapidly earned its reputation as a state-of-the-art open-source LLM, offers 7b, 13b, and 70b versions. Given computational constraints, we chose the 7b version of each model in this work.
Notably, models like Bard\cite{bard}, and GPT4\cite{openai2023gpt4}, which currently exhibit state-of-the-art performance and are widely used, were excluded from our study due to their closed-source nature and lack of fine-tuning support.

\begin{table*}[t]
\centering
\caption{Description of models evaluated and Fine-tuning Hyperparameters}
\label{tab:model}
\resizebox{\linewidth}{!}{%
\begin{tabular}{l|c|c|c|c|c|c|c|c|c}
\hline
\textbf{Model}         & \textbf{\#Param} & \textbf{Layers} & \textbf{Hidden Size} & \textbf{\begin{tabular}[c]{@{}c@{}}Content\\ Length\end{tabular}} & \textbf{\begin{tabular}[c]{@{}c@{}}Learning\\ Rate\end{tabular}} & \textbf{$\beta_1$}   & \textbf{$\beta_2$}      & \textbf{\begin{tabular}[c]{@{}c@{}}\# of Trainable\\ Parameters\end{tabular}} & \textbf{\begin{tabular}[c]{@{}c@{}}\# of Non-Trainable\\ Parameters\end{tabular}} \\ \hline
StableLM-Base-Alpha-7b & 7,868,755,968    & 16              & 6144                 & 4096                                                              & 0.00016                                                          & \multirow{3}{*}{0.9} & \multirow{3}{*}{0.9999} & 3,136,672                                                                     & 7,868,350,464                                                                     \\ \cline{1-6} \cline{9-10} 
Falcon-7b              & 6,921,720,704    & 32              & 4544                 & 2048                                                              & 0.0006                                                           &                      &                         & 3,839,186                                                                     & 7,216,889,856                                                                     \\ \cline{1-6} \cline{9-10} 
Llama-2 7b             & 6,738,415,616    & 32              & 4096                 & 4096                                                              & 0.0003                                                           &                      &                         & 4,279,744                                                                     & 6,738,149,376                                                                     \\ \hline
\end{tabular}%
}
\end{table*}

\subsubsection{Fine-Tuning Process}
The selected LLMs were fine-tuned by utilizing the LLaMA-Adapter V2~\cite{gao2023llamaadapter}, a method by the introduction of learnable adaptation prompts within specific layers in the Transformer model, thereby unlocking additional trainable parameters. These parameters significantly enhance the model's learning capability and can more efficiently adapt to the specific tasks. The fine-tuning process is elaborated in Algorithm~\ref{alg:fine-tuning}.
Table~\ref{tab:model} also provides a detailed account of the number of trainable and non-trainable parameters for each model. Applying Adapter V2 ensures the model's adaptability without significantly increasing the size or computational demand, making the fine-tuning more efficient and effective.

\begin{algorithm}
\caption{Fine-tuning Training Process with Adapter v2}
\label{alg:fine-tuning}
\begin{algorithmic}
\Procedure{Train}{$model, optimizer, train\_data$}
    \State \textbf{Initialize} training status
    \State \textbf{AddAdapterParametersToLinearLayers}($model$) 
    \State \textbf{MarkOnlyAdapterAsTrainable}($model$) 
    \For{each iteration in max\_iterations}
        \State Update learning rate
        \State $input\_ids, targets \gets$ \textbf{GetBatch}($train\_data$)
        \State $logits, loss \gets$ \textbf{ComputeLoss}($model, input\_ids, targets$)
        \State \textbf{Backward and Optimize}($loss, optimizer$) 
        \If{iteration $\bmod$ eval\_interval == 0}
            \State \textbf{Validate}($model, val\_data$)
            \State \textbf{SaveModel}($model$) \Comment{Save with adapter parameters}
        \EndIf
    \EndFor
\EndProcedure
\end{algorithmic}
\end{algorithm}


\subsection{Downstream Tasks and Evaluation Metrics}
This section elaborates on two downstream tasks, bug localization and repair, and the metrics for evaluating their performance.

\subsubsection{Bug Localization}

\projectname{} constructed a bug localization test set composed of original design code and code removed during repair (regarded as defective parts) sourced from the validation set. This code corresponds to hardware designs scripted in (System)Verilog. Under the prompt \textit{``Could you identify the possible bug inside the design?''}, we input the original hardware design into our fine-tuned LLM, which outputs a potentially defective statement.

\subsubsection{Bug Repair}

In \projectname{}, our focus is on leveraging the capabilities of fine-tuned LLMs for identifying and repairing bugs. We extract the original and repaired design from the validation set to serve as our golden model for comparison. To assess the model's repair function, a prompt: ``\textit{Could you fix the possible bug inside the design?}'' is provided for all fine-tuned models. Given the original design as input, the model could output a refined or corrected version. The output is subsequently compared to the repaired version from manual repair via ROUGE-N F1 score\cite{lin2004rouge}. ROUGE-N refers to the direct n-gram overlap between a  prediction and a reference word. And ROUGE F1 score derived from n-gram precision and recall.


\section{Implementation and Demonstration}\label{sec:expr}

\subsection{Experiment Setup}

The fine-tuning process was carried out on a server running Ubuntu 20.04.6 LTS, equipped with an Intel(R) Xeon(R) Silver 4314 CPU (2.40 GHz, 64 cores), 251 GB of memory, and dual NVIDIA A100 80 GB graphics cards. We utilized PyTorch's Fully Sharded Data Parallel API~\cite{zhao2023pytorch} for parallel acceleration, with the AdamW optimizer~\cite{loshchilov2019decoupled} used for weight updates and loss minimization. As shown in Table~\ref{tab:model}, different learning rates were applied based on each model's recommendation. These hyperparameters represent typical values found in the literature \cite{gpt-neox-library,falcon40b,touvron2023llama}. We utilized bf16-mixed precision to expedite training and reduce the model size on the A100s.

\subsection{Fine-tuned LLMs Performance Evaluation and Comparisons}



Eliminating comments and indents from the training dataset, undertaken to conserve content length, has reduced Rouge scores. Table \ref{tab:comparison-finetuned-models} presents the automatic evaluation of \projectname{} for correctly locating commit modifications on the validation set. Specifically, F1 scores for four metrics are applied: Rouge-1, Rouge-2, Rouge-L, and Rouge-W(weight-factor:1.2). Additionally, we include the organizations, Repo name, and the corresponding Git Commit SHA.\textbf{ The result for LLMs without fine-tuned falls from 0 to 0.02. Due to its lack of significance, it is not compared in the table.} Red numbers highlight the best model in the case.

We observe that the three models demonstrate proficiency in identifying potential bugs. When evaluated collectively, they can deliver highly accurate results. However, the stability of their outputs varies significantly. Distinct hardware designs elicit vastly different performance metrics from these models. Despite the marginal differences in size, we attribute this inconsistency to two factors:
\begin{enumerate}[left=0pt]
    \item Attention discrepancy: the attention heads' quantity and parameters differ among the models. The potential misalignment between attention parameters designed for natural language processing and the nuances of hardware design source code may introduce instability.
    \item Dataset size: More data to ensure consistent performance.
\end{enumerate}

In bug repair, LLama2 outperforms Falcon and StableLM. Ensuring that the models' proficiency in finding bugs surpasses their competence in fixing them is essential. Bug repair often demands more extensive knowledge compared to mere localization. We have noticed that models tend to prematurely terminate their generation before providing a comprehensive repaired hardware design. The model's cessation mechanism, ruled by predefined probability, implies that our models encounter confusion when tasked with more extended content generation. 

LLMs are designed to generalize across their training data, drawing upon vast amounts of information rather than focusing on specific content. Given this broad learning approach, relying solely on metrics like Rouge scores, which evaluate the overlap between reference and generated summaries, might not capture LLMs' full capabilities and nuances. Therefore, we provide two examples to demonstrate the effectiveness of \projectname{} and compare its performance with that of ChatGPT, BARD, and the original model without fine-tuning. The selection of these two instances aims to demonstrate the model's performance under different circumstances as comprehensively as possible, thus enabling a more accurate assessment of its capabilities.

\begin{table*}[t]
\centering
\caption{Comparative Evaluation of Different Fine-Tuned Models for Hardware Design Downstream Tasks}
\label{tab:comparison-finetuned-models}
\resizebox{\textwidth}{!}{%
\begin{tabular}{c|c|c|c|l|c|c|c|c}
\hline
\textbf{Downstream Task} & \textbf{Organization} & \textbf{Repository Name} & \textbf{Git Commit SHA} & \multicolumn{1}{c|}{\textbf{Model}} & \textbf{ROUGE-1 F1 Score} & \textbf{ROUGE-2 F1 Score} & \textbf{ROUGE-L F1 Score} & \textbf{ROUGE-W F1 Score} \\ \hline
 &  &  &  & Falcon-7B & 0.666666667 & 0.626087 & 0.71896 & 0.417834 \\ \cline{5-9} 
 &  &  &  & Llama2-7B & 0.626086957 & 0.619469 & 0.683108 & 0.399298 \\ \cline{5-9} 
 & \multirow{-3}{*}{\begin{tabular}[c]{@{}c@{}}OpenHW \\ Group\end{tabular}} & \multirow{-3}{*}{\begin{tabular}[c]{@{}c@{}}CORE V\\ MCU\end{tabular}} & \multirow{-3}{*}{\begin{tabular}[c]{@{}c@{}}580275ce67d3c\\ 8fa92faeff082\\ 8b0f4b335c8bfe\end{tabular}} & StableLM-base-alpha-7b & {\color[HTML]{FE0000} 0.677966102} & {\color[HTML]{FE0000} 0.672414} & {\color[HTML]{FE0000} 0.728704} & {\color[HTML]{FE0000} 0.428563} \\ \cline{2-9} 
 &  &  &  & Falcon-7B & 0.369863014 & 0.277778 & 0.414738 & 0.214051 \\ \cline{5-9} 
 &  &  &  & Llama2-7B & {\color[HTML]{FE0000} 0.575163399} & {\color[HTML]{FE0000} 0.556291} & {\color[HTML]{FE0000} 0.639046} & {\color[HTML]{FE0000} 0.370246} \\ \cline{5-9} 
 &  &  & \multirow{-3}{*}{\begin{tabular}[c]{@{}c@{}}fb115220b0c85\\ 70ee773f4d609\\ 501f28bd72e600\end{tabular}} & StableLM-base-alpha-7b & 0.438356164 & 0.388889 & 0.512094 & 0.305373 \\ \cline{4-9} 
 &  &  &  & Falcon-7B & 0.436363636 & 0.333333 & 0.506878 & 0.26954 \\ \cline{5-9} 
 &  &  &  & Llama2-7B & {\color[HTML]{FE0000} 0.805555556} & {\color[HTML]{FE0000} 0.8} & {\color[HTML]{FE0000} 0.837331} & {\color[HTML]{FE0000} 0.50795} \\ \cline{5-9} 
 &  &  & \multirow{-3}{*}{\begin{tabular}[c]{@{}c@{}}d914eb9becfd1\\ 5cdec953072ec\\ 6d74be2b6054d6\end{tabular}} & StableLM-base-alpha-7b & 0.540540541 & 0.458716 & 0.588562 & 0.333338 \\ \cline{4-9} 
 &  &  &  & Falcon-7B & 0.269967645 & 0.217939 & 0.332595 & 0.17633 \\ \cline{5-9} 
 &  &  &  & Llama2-7B & {\color[HTML]{FE0000} 0.705882353} & {\color[HTML]{FE0000} 0.68} & {\color[HTML]{FE0000} 0.75268} & {\color[HTML]{FE0000} 0.463655} \\ \cline{5-9} 
 &  &  & \multirow{-3}{*}{\begin{tabular}[c]{@{}c@{}}03c95e8d9e6ac\\ d1da33069134d\\ 4888ad4f759b8c\end{tabular}} & StableLM-base-alpha-7b & 0.505354752 & 0.499423 & 0.575968 & 0.338396 \\ \cline{4-9} 
 &  &  &  & Falcon-7B & 0.339896188 & 0.22258 & 0.405916 & 0.221039 \\ \cline{5-9} 
 &  &  &  & Llama2-7B & {\color[HTML]{FE0000} 0.49704142} & {\color[HTML]{FE0000} 0.491018} & {\color[HTML]{FE0000} 0.569061} & {\color[HTML]{FE0000} 0.324888} \\ \cline{5-9} 
\multirow{-15}{*}{\begin{tabular}[c]{@{}c@{}}Fix the Possible \\ Bug inside the design\end{tabular}} & \multirow{-12}{*}{lowRISC} & \multirow{-12}{*}{OpenTitan} & \multirow{-3}{*}{\begin{tabular}[c]{@{}c@{}}03fbb03f78db0\\ e0565a359cc68\\ 32f88a41d69dbb\end{tabular}} & StableLM-base-alpha-7b & 0.479687983 & 0.417485 & 0.551211 & 0.313176 \\ \hline
 &  &  &  & Falcon-7B & {\color[HTML]{FE0000} 1} & {\color[HTML]{FE0000} 1} & {\color[HTML]{FE0000} 1} & {\color[HTML]{FE0000} 0.890579} \\ \cline{5-9} 
 &  &  &  & Llama2-7B & 0.181818182 & 0 & 0.245251 & 0.147212 \\ \cline{5-9} 
 & \multirow{-3}{*}{\begin{tabular}[c]{@{}c@{}}OpenHW \\ Group\end{tabular}} & \multirow{-3}{*}{\begin{tabular}[c]{@{}c@{}}CORE V\\ MCU\end{tabular}} & \multirow{-3}{*}{\begin{tabular}[c]{@{}c@{}}580275ce67d3c\\ 8fa92faeff082\\ 8b0f4b335c8bfe\end{tabular}} & StableLM-base-alpha-7b & 0.139534884 & 0 & 0.207069 & 0.084116 \\ \cline{2-9} 
 &  &  &  & Falcon-7B & 0.487201989 & 0.398693 & 0.52284 & 0.326274 \\ \cline{5-9} 
 &  &  &  & Llama2-7B & {\color[HTML]{FE0000} 1} & {\color[HTML]{FE0000} 1} & {\color[HTML]{FE0000} 1} & {\color[HTML]{FE0000} 0.706368} \\ \cline{5-9} 
 &  &  & \multirow{-3}{*}{\begin{tabular}[c]{@{}c@{}}143df43e328c62\\ fa08ac5cb64d0\\ 404ccb2a8f0c9\end{tabular}} & StableLM-base-alpha-7b & 0.454203122 & 0.435464 & 0.495669 & 0.301564 \\ \cline{4-9} 
 &  &  &  & Falcon-7B & 0.2 & 0.058824 & 0.261591 & 0.134438 \\ \cline{5-9} 
 &  &  &  & Llama2-7B & 0.315789474 & 0.290909 & 0.385101 & 0.261872 \\ \cline{5-9} 
 &  &  & \multirow{-3}{*}{\begin{tabular}[c]{@{}c@{}}cfcfde74dc778\\ 81ca47870f706\\ 320cf042cd26f9\end{tabular}} & StableLM-base-alpha-7b & {\color[HTML]{FE0000} 0.615384615} & {\color[HTML]{FE0000} 0.447368} & {\color[HTML]{FE0000} 0.667372} & {\color[HTML]{FE0000} 0.392397} \\ \cline{4-9} 
 &  &  &  & Falcon-7B & {\color[HTML]{FE0000} 0.253229974} & {\color[HTML]{FE0000} 0.166667} & 0.289436 & {\color[HTML]{FE0000} 0.221765} \\ \cline{5-9} 
 &  &  &  & Llama2-7B & 0.222222222 & 0 & {\color[HTML]{FE0000} 0.292272} & 0.162201 \\ \cline{5-9} 
 &  &  & \multirow{-3}{*}{\begin{tabular}[c]{@{}c@{}}03fbb03f78db0\\ e0565a359cc68\\ 32f88a41d69dbb\end{tabular}} & StableLM-base-alpha-7b & 0.034199134 & 0 & 0.060415 & 0.024152 \\ \cline{4-9} 
 &  &  &  & Falcon-7B & 0.391562314 & 0.311727 & 0.441126 & 0.274217 \\ \cline{5-9} 
 &  &  &  & Llama2-7B & {\color[HTML]{FE0000} 0.833333333} & {\color[HTML]{FE0000} 0.818182} & {\color[HTML]{FE0000} 0.860715} & {\color[HTML]{FE0000} 0.624436} \\ \cline{5-9} 
 &  &  & \multirow{-3}{*}{\begin{tabular}[c]{@{}c@{}}9c922ae8c623c2\\ e4d4ef71ceff1\\ 506a2e7170fcd\end{tabular}} & StableLM-base-alpha-7b & 0.213178295 & 0.190432 & 0.243214 & 0.181597 \\ \cline{4-9} 
 &  &  &  & Falcon-7B & {\color[HTML]{FE0000} 0.5} & {\color[HTML]{FE0000} 0.263158} & {\color[HTML]{FE0000} 0.571747} & {\color[HTML]{FE0000} 0.271375} \\ \cline{5-9} 
 &  &  &  & Llama2-7B & 0.045454545 & 0 & 0.082274 & 0.031019 \\ \cline{5-9} 
\multirow{-18}{*}{\begin{tabular}[c]{@{}c@{}}Find the possible\\ bug inside the input\end{tabular}} & \multirow{-15}{*}{lowRISC} & \multirow{-15}{*}{OpenTitan} & \multirow{-3}{*}{\begin{tabular}[c]{@{}c@{}}4172d4d7c2a13\\ a16e69421e16e\\ 2fa0bded39ea0f\end{tabular}} & StableLM-base-alpha-7b & 0 & 0 & 0 & 0 \\ \hline
\end{tabular}%
}
\end{table*}
\subsection{Results and Analysis: Bug Localization}


In the Bug Localization task, we choose a modification in OpenTitan. The modified code at \cite{debug1} plays a vital role in hardware verification, defining a JTAG DPI module for interactions with the JTAG interface. In a recent modification, the module was updated by Section 6.14 of the IEEE Standard for SystemVerilog (1800-2017) \cite{8299595}, which states, \textit{``Chandles should always be initialized to the value null''}, equating to 0 in C. In the original design code, \lstinline[style=vcode1]{ctx=0} was revised to \lstinline[style=vcode1]{ctx=null}. This commit addressed issues in three hardware designs, incorporating two into the training set and allocating one to the validation set.

All our models correctly locate the change by outputting the entire line content, disregarding indentation. In contrast, the base model, which had not been fine-tuned, refused to accept this bug-locating inquiry.
BARD and ChatGPT offered insightful but differing perspectives. ChatGPT focused on handling the \lstinline[style=vcode1]{rst_ni} reset signal in the \lstinline[style=vcode1]{always_ff} block, highlighting a potential issue with the structure of the block itself. BARD, meanwhile, pointed out the lack of initialization for the \lstinline[style=vcode1]{ctx} variable in the \lstinline[style=vcode1]{always_ff} block, which could lead to undefined behavior when \lstinline[style=vcode1]{jtagdpi_tick} the function is invoked.
Both models, however, failed to directly identify the crux of the problem, that is, the incorrect initialization of \lstinline[style=vcode1]|ctx| that violates the guideline of the 1800-2017 IEEE Standard for SystemVerilog.



\subsection{Results and Analysis: Bug Repair}

We show a case study of preparing a bug from the Base proxy class for all security countermeasure interfaces in OpenTitan \cite{opentitan}.

In the Listing \ref{lst:sec_cm_base_if_proxy_orinal}, lines prefixed with the symbol '-' denote the original code, whereas those prefixed with '+' indicate the revised segments.
The original hardware design employs a base proxy class as the cornerstone for all security countermeasure interfaces. Notably, the \lstinline[style=vcode1]{inject_fault()} and \lstinline[style=vcode1]{restore_fault()} pure virtual tasks are initially static, meaning all variables are instantiated on their first call and destroyed by the simulation's end. A design flaw is that if a task were invoked more than once, every call would utilize the same variable, exposing the system to a significant security vulnerability.

Recognizing this vulnerability, lowRISC introduces the \lstinline[style=vcode1]{automatic} keyword to guarantee that every task or function call would allocate new storage space for the associated variables. Consequently, each function or task call now reserves new storage for the variables, which is then relinquished upon the function or task's completion. It ensures each task invocation operates with an independent variable instance, preventing potential complications from state sharing.

\begin{lstlisting}[style={vcode},breaklines=true,label={lst:sec_cm_base_if_proxy_orinal},caption={Opentitan Proxy Class Adjustments}]
 // Copyright lowRISC contributors.
 // Licensed under the Apache License, Version 2.0, see LICENSE for details.
 // SPDX-License-Identifier: Apache-2.0

 // This is the base proxy class for all the sec_cm interfaces.
 virtual class sec_cm_base_if_proxy extends uvm_object;
   sec_cm_type_e sec_cm_type;
   string path; 

   `uvm_object_new
 
-  pure virtual task inject_fault();
-  pure virtual task restore_fault();
+  pure virtual task automatic inject_fault();
+  pure virtual task automatic restore_fault();
 endclass
\end{lstlisting}
 





This bug-fixing effort involved modifications across multiple files, including three in the training set, and demonstrated cases from the validation dataset. The fine-tuned Falcon-7B and LLama2 7B model learned the pattern and correctly applied the fix in Listing \ref{lst:sec_cm_base_if_proxy_Finetuned_Falcon_7B}. 
\begin{lstlisting}[style={vcode},label={lst:sec_cm_base_if_proxy_Finetuned_Falcon_7B},caption={Response from Finetuned Falcon 7B and LLama2 7B with prompt: ``Fix the possible BUG inside the given hardware design.''}]
virtual class sec_cm_base_if_proxy extends uvm_object;
  sec_cm_type_e sec_cm_type;
  string path;
  `uvm_object_new
  pure virtual task automatic inject_fault();
  pure virtual task automatic restore_fault();
endclass
\end{lstlisting}

The base Falcon 7B model's performance, which has yet to be fine-tuned, is substantially different. As shown in Listing \ref{lst:sec_cm_base_if_proxy_falcon}, the model's response illustrates its lack of specialized hardware design domain-specific training. The model correctly recognized its limitations. Despite not providing a direct solution to the issue, it offered valuable advice—underlining the necessity of professional assistance when dealing with complex hardware problems.

\begin{lstlisting}[stringstyle=\color{string}, basicstyle=\scriptsize\ttfamily, captionpos=b, numbers=none,	frame=single, showstringspaces=false,tabsize=2,breaklines=true,label={lst:sec_cm_base_if_proxy_falcon},caption={Response from base Falcon with prompt: ``Fix the possible BUG inside the given hardware design.''}]
I'm sorry, I cannot provide a solution to fix a possible BUG inside the given hardware design. It requires knowledge of hardware design, software development, and debugging techniques. It is recommended to consult with experienced hardware and software engineers or seek professional help.
\end{lstlisting}

For the Llama2 model, we opted for the llama2 70b-chat version, which boasts an average performance that is $22\%$ superior to the 7B baseline model\cite{Open-LLM-Leaderboard-Report-2023} for comparison. In Listing \ref{lst:sec_cm_base_if_proxy_LLama2}, it chose not to respond and recommended seeking advice from a professional.
\begin{lstlisting}[stringstyle=\color{string}, basicstyle=\scriptsize\ttfamily, captionpos=b, numbers=none,	frame=single, showstringspaces=false,tabsize=2,breaklines=true,label={lst:sec_cm_base_if_proxy_LLama2},caption={Response from LLama2-70b-chat with prompt: ``Fix the possible BUG inside the given hardware design.''}]
I cannot provide an answer to this question.
//..............Omitted section for clarity.............//
Instead, I suggest that you seek assistance from a qualified hardware engineer or software developer who can evaluate the entirety of your code and provide appropriate guidance on how to address any issues. 
//..............Omitted section for clarity.............//
\end{lstlisting}

Concurrently, ChatGPT4(July 19, 2023 Version) provides a high-level perspective in its responses in Listing \ref{lst:sec_cm_base_if_proxy_ChatGPT4}, emphasizing the importance of method implementations in subclasses and the significance of hardware design or simulation testing. However, it did not directly identify the crux of the problem—namely, that the \lstinline[style=vcode1]{inject_fault()} and \lstinline[style=vcode1]{restore_fault()} pure virtual tasks might share states when called multiple times, potentially causing errors. This may stem from GPT4 not receiving training in the hardware design domain and may be incapable of responding to issues.
\vspace{2pt}
\begin{lstlisting}[stringstyle=\color{string}, basicstyle=\scriptsize\ttfamily, captionpos=b, numbers=none,	frame=single, showstringspaces=false,tabsize=2,breaklines=true,label={lst:sec_cm_base_if_proxy_ChatGPT4},caption={Response from ChatGPT4 (July 19, 2023 Version) with prompt: ``Fix the possible BUG inside the given hardware design.''}]
//..............Omitted section for clarity.............//
The code you provided is a class definition with some virtual tasks, which are intended to be overridden by subclasses. There's nothing inherently wrong with the code you've provided. However, it's worth noting that bugs in hardware designs or simulations typically don't originate from such high-level descriptions or interface definitions. Instead, they tend to stem from the implementation of these interfaces and the specific logic of the design. If you suspect that there's a bug in your hardware design, I would recommend inspecting the implementation of these methods in the child classes, and checking the simulation testbenches or the hardware design itself.
//..............Omitted section for clarity.............//
\end{lstlisting}

While BARD (2023.07.13 version), as indicated in Listing \ref{lst:sec_cm_base_if_proxy_BARD}, can pinpoint potential issues in the code, especially the declaration of these two tasks as pure virtual functions which could result in undefined behavior if not defined in subclasses, this isn't the primary concern. Its proposed solution (changing pure virtual functions to virtual functions) cannot address the problem. This might also be due to BARD not receiving hardware design domain-specific training, thus potentially lacking the capability to respond to such problems.
\begin{lstlisting}[stringstyle=\color{string}, basicstyle=\scriptsize\ttfamily, captionpos=b, numbers=none,	frame=single, showstringspaces=false,tabsize=2,breaklines=true,label={lst:sec_cm_base_if_proxy_BARD},caption={Response from BARD (2023.07.13 version) with prompt: ``Fix the possible BUG inside the given hardware design.''}]
The possible bug in the given hardware design is that the inject_fault() and restore_fault() tasks are declared as pure virtual. This means that they must be overridden by any subclass of sec_cm_base_if_proxy. However, if a subclass does not override these tasks, then they will be left undefined, which could lead to undefined behavior.To fix this bug, the inject_fault() and restore_fault() tasks should be declared as virtual instead of pure virtual. This will allow subclasses to override these tasks without having to define them.
//..............Omitted section for clarity.............//
\end{lstlisting}



\section{Conclusion}\label{sec:con}

This paper presents \projectname{}, a groundbreaking framework that leverages domain-specific LLMs for hardware debugging. 
Our evaluation shows our method's effectiveness in accurately identifying and correcting defects, offering a new perspective on automating the quality control process in hardware design. 
Moreover, our findings corroborate that, despite emerging architectures, there needs to be a subtle but not pronounced performance variance in models with similar parameter sizes post-fine-tuning. This necessitates data to train LLM and acquire a computational platform capable of supporting larger-parameter LLMs. Such requirements introduce new challenges when applying fine-tuned LLMs for hardware debugging and security.


\normalem
\bibliographystyle{IEEEtran}
\bibliography{Xiaolong}
\end{document}